\documentclass[a4paper,11pt]{article}
\usepackage[]{fontenc}
\usepackage[utf8]{inputenc}
\usepackage[]{graphicx}

\usepackage{natbib}

\usepackage{fancyhdr}
\title{\bf MULTIVARIATE EVOLUTIONARY ANALYSES IN ASTROPHYSICS}
\author{\bf Didier Fraix-Burnet \\
\small Université Joseph Fourier - Grenoble 1 / CNRS \\
\small Laboratoire d'Astrophysique de Grenoble (LAOG) UMR 5571 \\
\small BP 53, F-38041 GRENOBLE Cedex 09, France \\
\small Email: fraix@obs.ujf-grenoble.fr}
\date{September 2010}

\begin{document}

\maketitle
\thispagestyle{fancy}

\fancyhead[L]{Astronomical Data Analysis, 6th conf., 3-7 May 2010, Monastir, Tunisia}
\fancyhead[R]{}

\begin{abstract}
The large amount of data on galaxies, up to higher and higher redshifts, asks for sophisticated
statistical approaches to build adequate classifications. Multivariate cluster analyses, that 
compare objects for their global similarities, are still confidential in astrophysics, probably 
because their results are somewhat difficult to interpret. We believe that the missing key is the 
unavoidable characteristics in our Universe: evolution. Our approach, known as Astrocladistics, 
is based on the evolutionary nature of both galaxies and their properties. It gathers objects 
according to their “histories” and establishes an evolutionary scenario among groups of objects. 
In this presentation, I show two recent results on globular clusters and earlytype galaxies to 
illustrate how the evolutionary concepts of Astrocladistics can also be useful for multivariate 
analyses such as K-means Cluster Analysis.

\end{abstract}

\section{Introduction}

We are now able to study galaxies in great detail, identifying individual stars, gas and dust clouds, as well as different stellar populations. Imagery brings very fine structural details, and spectroscopy provides the kinematical, physical and chemical conditions of the observed entities at different locations within a galaxy. For more distant objects, information is scarcer, but deep systematic sky surveys gather spectra for millions of galaxies at various redshifts.
The amount of data on galaxies, their number, their diversity, their complexity and that of their evolution, suggest that they should be envisaged as a population or an ensemble of populations. This implies the use of the appropriate statistical tools. 

Like paleontologists, we observe objects from the distant past (galaxies at high redshift), and like evolutionary biologists, we want to understand their relationships with nearby galaxies, like our own Milky Way. Consequently, a "galactogenesis" can be advantageously approached by considering phylogenetics methods.

\section{Why be multivariate?}

The description of a given galaxy requires many observables, most of them derived from a spectrum. Usual classifications, often inspired by the Hubble tuning fork, use only a very few properties. Even if bivariate plots or correlations are clear and useful, they are incomplete. Worse, they are merely the projection onto a 2-D diagram of a multivariate parameter space. This projection is generally expected to increase the dispersion of the plot. Anyhow, it is difficult to represent many data with only bivariate plots, and any classification necessarily requires an arbitrary binning of one or several parameters.

Multivariate analyses are still not much used in astrophysics. One basic tool, the Principal Component Analysis, is relatively well-known \citep[e.g.][]{cabanac2002, RecioBlanco2006}, but this is not a clustering tool in itself. A very few attempts to apply multivariate clustering methods have been made very recently \citep{Chattopadhyay2006,Chattopadhyay2007,ChattopadhyayGRB2007, Chattopadhyay2008,Chattopadhyay2009a,Chattopadhyay2009b,FDC09,Fraix2010}. Sophisticated statistical tools are used in some areas of astrophysics and are developing steadily, but multivariate analysis and clustering techniques have not much penetrated the community. It is true that the interpretation of the results are not always easy.

\section{Why be evolutive?}

Evolution, an unavoidable fact, is also not correctly taken into account in most classification methods. By mixing together objects at different stages of evolution, most of the physical significance and usefulness of a classification is lost. In practice, the evolution of galaxies is often limited to the evolution of the properties of the entire population as a function of redshift \citep{Bell2005}. Since environment (the expanding Universe) and galaxy properties are so much intricate, this kind of study is relevant to a first approximation. However, recent observations have revealed that galaxies of all kinds do not evolve perfectly in parallel, as illustrated for instance by the so-called downsizing effect which shows that large galaxies formed their stars earlier than small ones \citep[e.g.][]{Neistein2006}. New observational instruments now bring multivariate information at different stages of evolution, and in various evolutive environments. In this multivariate context, we believe that the notion of "evolution", easy to understand for a single parameter, is advantageously replaced by "diversification".

The transformation of galaxies is a complex process \citep{jc1,jc2} that cannot be disentangled with only a very few observables. For instance, the elliptical shape of galaxy can be obtained through the monolithic collapse of a big cloud of gas, or by big mergers. To find which process has shaped a given galaxy, many observables are required. Only a multivariate and evolutive analysis can distinguish different histories.

\section{Classification, complexity, evolution}

Multivariate clustering methods compare objects with a given measure and then gather them according to a proximity criterion. There are two main classes. Firstly, distance analyses are based on the overall similarity derived from the values of the parameters describing the objects. The choice of the most adequate distance measure for the data under study is not unique and remains difficult to justify a priori. The way objects are subsequently grouped together is also not uniquely defined. Secondly, methods based on characters (a trait, a descriptor, an observable, or a property, that can be given at least two states characterizing the evolutionary stages of the object for that character) compare objects in their evolutionary relationships \citep{Wiley1991}. Here, the “distance" is an evolutionary cost simply measured by the number of changes of the parameter values (or character states). Groupings are then made on the basis of shared or inherited characteristics, and are most conveniently represented on an evolutionary tree.

Character-based methods like cladistics are better suited to the study of complex objects in evolution, even though the relative evolutionary costs of the different characters is not easy to assess. Distance-based methods are generally faster and often produce comparable results, but the overall similarity is not always adequate to compare evolving objects. In any case, one has to choose a multivariate method, and the results are generally somewhat different depending on this choice \citep{buchanan2008}. However, the main goal is to reveal a hidden structure in the data sample, and the relevance of the method is mainly provided by the interpretation and usefulness of the result. 

We must note that taking all available parameters blindly can kill the multivariate and evolutive analysis. One dangerous component is a hidden correlation, such as a size effect, that creates a redundancy. A less known caveat is due to spurious correlations, due to independant variables that vary as function of a non-necessarily obvious parameter. This is especially the case with the time or the stage of evolution. Two quantities can be totally unrelated but if they vary both with time in a more or less monotonic way, then they appear to be correlated. For instance, all photometric quantities for galaxies are affected by the stellar evolution. In such a case, a cladistic analysis yields a regular tree showing the regular stellar evolution \citep[e.g.][]{DFB06}.

Multivariate evolutionary classification in astrophysics has been pioneered by the author \citep{FCD06,jc2, FDC09, Fraix2010, DFB09}. Called astrocladistics, it is based on cladistics that is heavily developed in evolutionary biology. Astrocladistics has been first applied to galaxies \citep{jc1,Fraix2010} because they can be shown to follow a transmission with modification process when they are transformed through assembling, internal evolution, interaction, merger or stripping. For each transformation event, stars, gas and dust are transmitted to the new object with some modification of their properties. Cladistics has also been applied to globular clusters \citep{FDC09}, where interactions and mergers are probably rare. These are thus simpler stellar systems, even though we have firm evidence that internal evolution can create another generation of stars and that globular clusters can lose mass. Basically, the properties of a globular cluster strongly depend on the environment in which it formed (chemical composition and dynamics), and also on the internal evolution which includes at least the aging of its stellar populations. Since galaxies and globular clusters form in a very evolving environment (Universe, dark matter haloes, galaxy clusters, chemical and dynamical environment), the basic properties of different objects are related to each other by some evolutionary pattern. 
\begin{figure}
                   \includegraphics[width=11 true cm,angle=0]{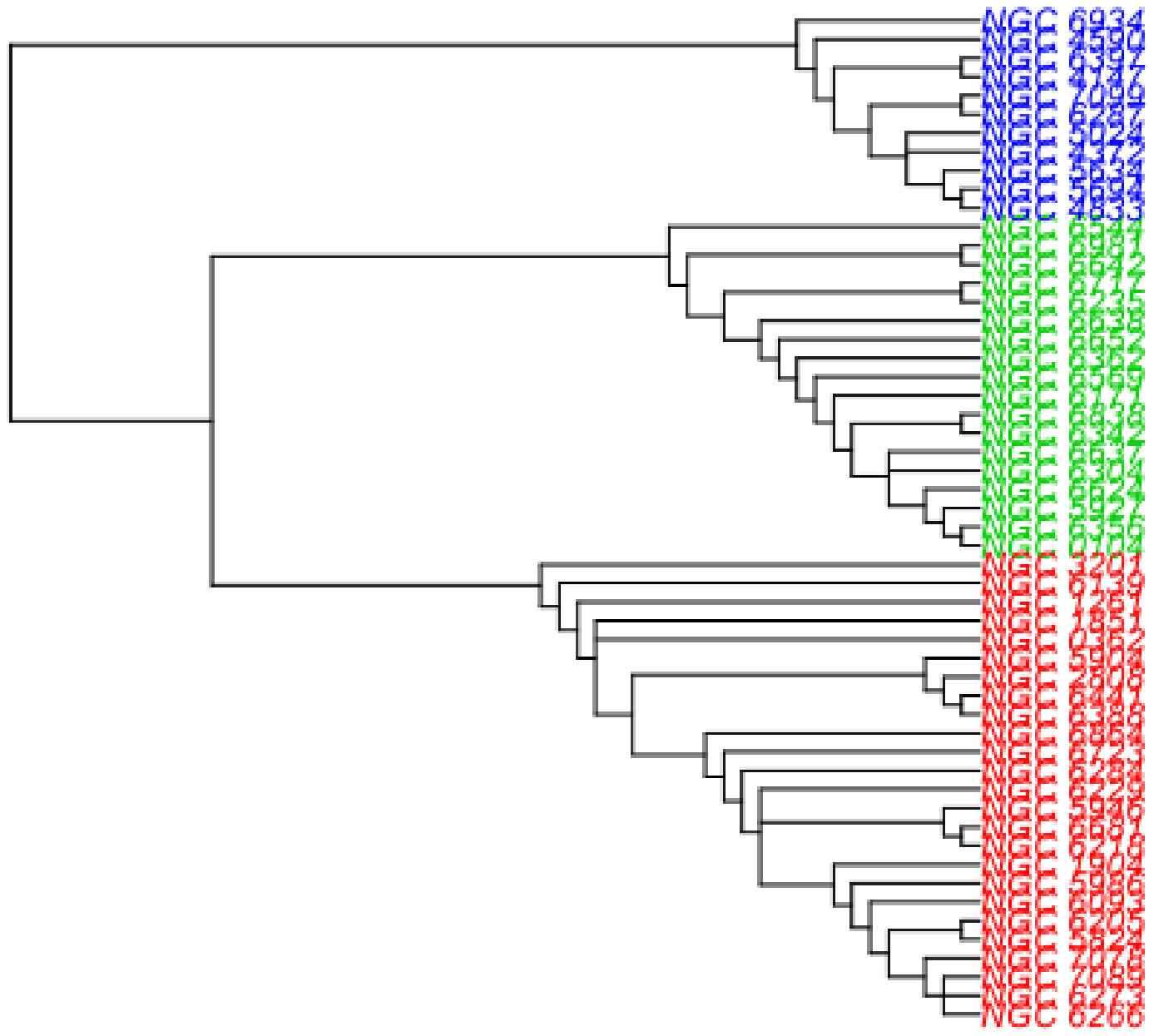}
                \includegraphics[width=11 true cm,angle=0]{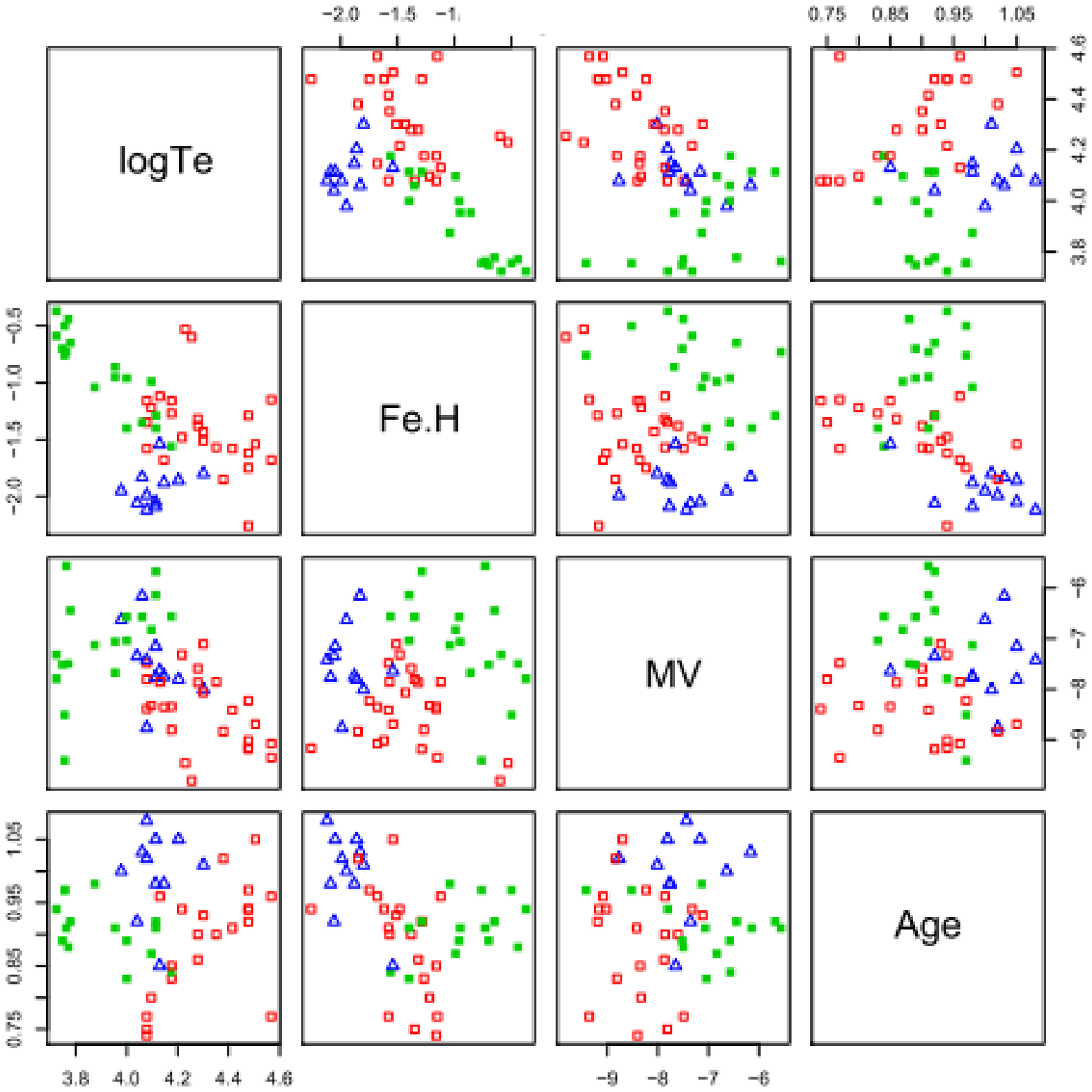}
\caption{\textit{Top}: cladogram obtained for the globular clusters of our Galaxy. \textit{Bottom}: projection of the partitioning on pair plots with the four parameters used for the cladistic analysis. From \citet{FDC09}.}
\label{figAG1}
\end{figure}

\section{A more pertinent physical interpretation}

An obvious difficulty for a physicist in general is to intepret the results of multivariate analyses using his models that mostly result from a set of equations and are more conveniently presented by curves on bivariate plots. Interestingly enough, these models are multivariate, especially in astrophysics, and the resolution of the set of equations yields a "population" of possible results often called a grid of models. As a result, some parameters are set to sensible values, and the corresponding models are then compared to some observables. These observables can also have been truncated by setting some other observables in order to simplify the information.  

It appears that we must here compare two populations, one of real objects and one of models, in a multivariate space. We show here two examples of multivariate (and evolutive) analyses of astrophysical objects showing that such approaches are both more direct, objective and physically pertinent.

Figure~\ref{figAG1} shows the cladogram obtained for globular clusters of our Galaxy  and the projection of the partitioning on pair plots for the four parameters used for the analysis:  logTe, that measures the temperature of stars that are at a specific point in their evolution,  Fe/H, MV that is the total visible intensity (magnitude) and roughly indicates the mass of the globular cluster, and Age that can be measured quite precisely because all stars of a given globular cluster are formed nearly at the same time. However Age is not an intrinsic property discriminating evolutionary groups since it evolves in the same way for all. But we gave it a half weight to arrange the objects within each group \citep{FDC09}. Three groups are identified. The first one (in blue) has on average the lower ratio Fe/H that measures the proportion of heavy atomic elements that are processed within stars. This group is consequently considered as more primitive. It is obivous that this partitioning would be impossible to obtain with only bivariate plots.

\begin{figure}
                   \includegraphics[width=11 true cm,angle=0]{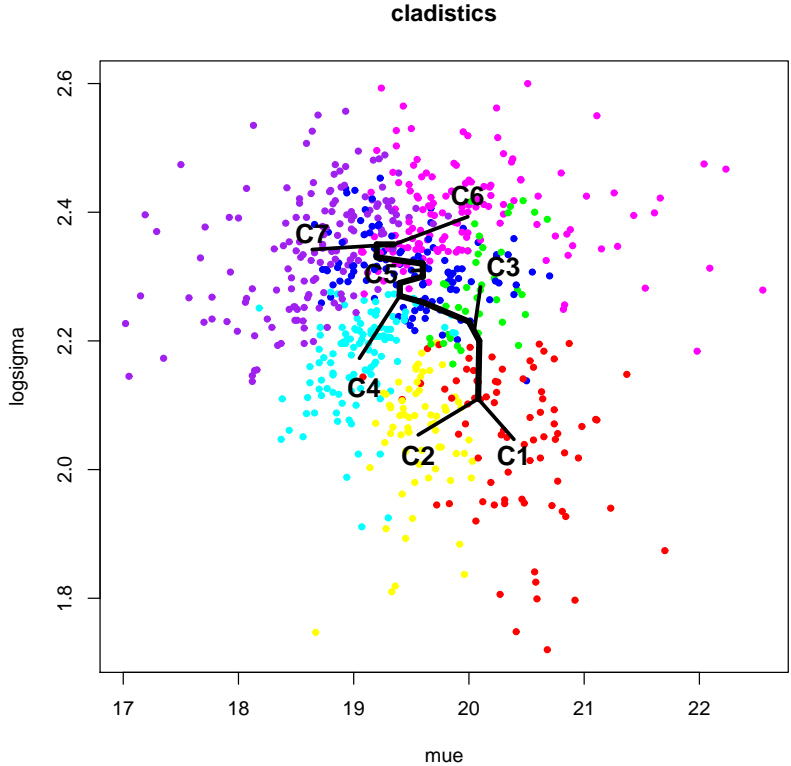}
 \caption{The fundamental plane of 699 galaxies, showing the partitioning and projection of the tree obtained by a cladistic analysis \citep{Fraix2010}.}
\label{PF1}
\end{figure} 
\begin{figure}
                   \includegraphics[width=11 true cm,angle=0]{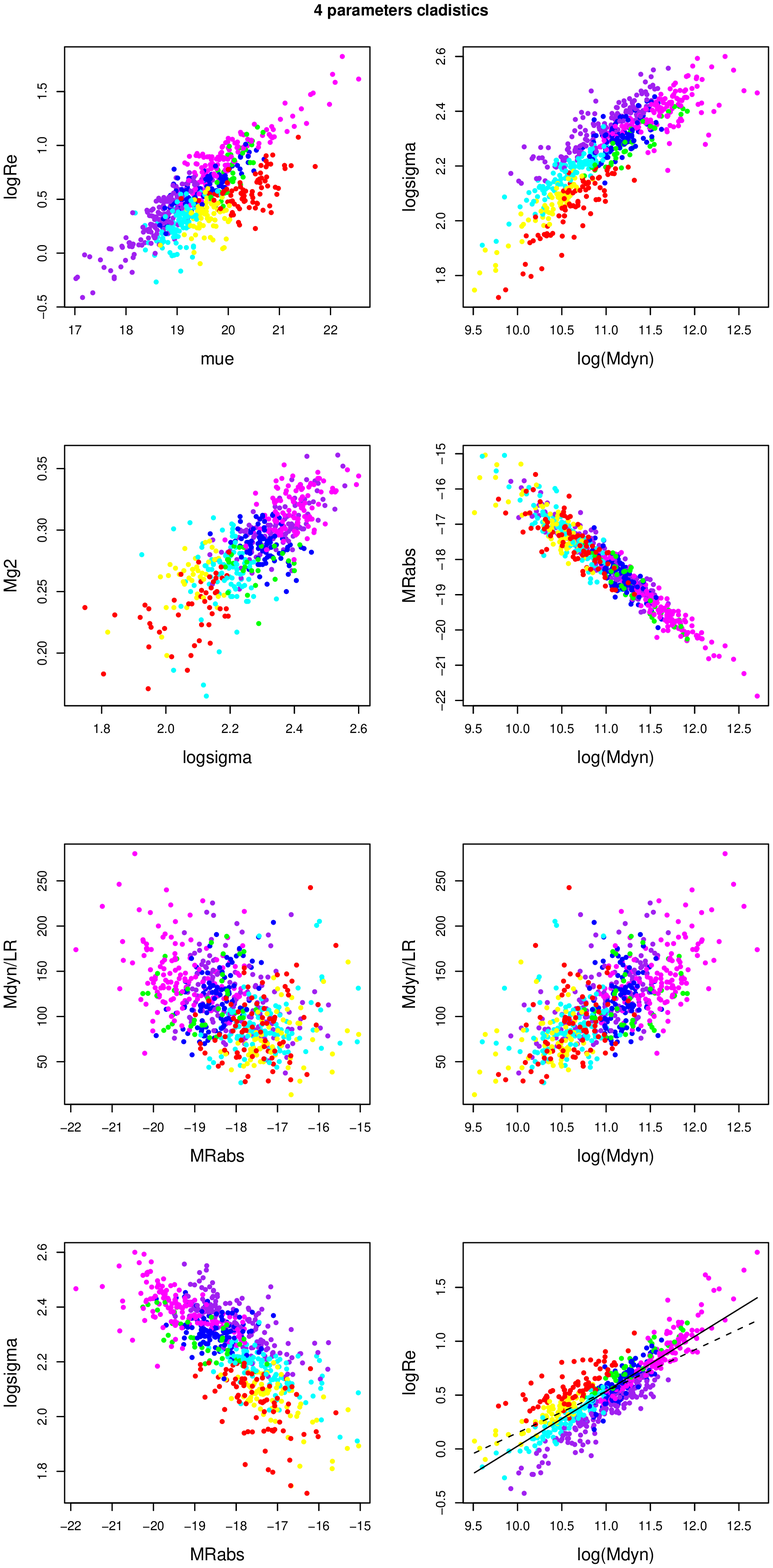}
 \caption{Cluster and cladistic analysis of the fundamental plane of early-type galaxies: bivariate plots showing how correlations differ for each group and for the whole sample. Note in particular the $Mg_2$ vs $\log\sigma$ plot revealing a spurious correlation between these two parameters \citep{Fraix2010}.}
\label{PF2}
\end{figure} 

Looking at other parameters (such as orbital elements, kinematics, more refined chemical abundances...) revealed clear characteristics that allowed us to infer that each group formed during a particular stage of the assembly history of our Galaxy. The blue group is the older one. It formed during the dissipationless collapse of the protogalaxy. They are located mainly in the outer halo. The red group belongs to the inner halo and the corresponding clusters formed at a later stage during the dissipational phase of Galactic collapse, which continued in the halo after the formation of the thick disc and its globular clusters. These clusters were very massive before "star evaporation" took place. The latter group (green) formed during an intermediate and relatively short period and comprises clusters of the disk of our Galaxy \citep[all details in][]{FDC09}.

Another example is given with the fundamental plane of early-type galaxies which is a lonog-known correlation between the central velocity dispersion $\sigma$, the surface brightness $\mu_e$ and the effective radius $r_e$. In addition, the metallicity, as measured with the $Mg_2$ index, plays a role and seems to be correlated with $\log\sigma$. We performed a K-means cluster analysis and a cladistic analysis in parallel \citep{Fraix2010}. The partionings are remarkably in agreement. We believe the reason is due to the careful choice of the parameters. For cladistics, they must be informative with respect to diversification, and should not be redundant or incompatible. This requirement is logically pertinent also for any cluster analysis. 

Cladistics provides in addition the evolutionary relationships between the groups. On Figure~\ref{PF1}, the tree is projected onto the $\log\sigma$ vs $\mu_e$ diagram on which the fundamental plane is seen essentialy face-on. Since galaxies are more complicated than globular clusters, the interpretation of the results and all relations between all possible parameters and within each group takes great advantage of numerical simulations. Here again, we are able to derive the probable history of each group of galaxies as well as their relative level of diversification, giving possible sequences of past transforming events such as mergers, accretions or sweeping \citep[for details, see][]{Fraix2010}.

A quite interesting finding is that most known correlations are different or even absent when we consider groups individually (Figure~\ref{PF2}). This proves that they have different evolution histories. Another noticeable fact is that the well-known correlation between Mg2 and $\log\sigma$ is indeed spurious, or historical. It is due to the fact that each parameter changes with the level of diversification as clearly shown by the placement of the groups (see Figure~\ref{PF2}).

\section{Conclusion}

Undoubtly, the study of galaxies now requires multivariate statistical treatments. Evolution must also be taken into account and the concept of populations seems appropriate and points to the use of methodologies developed elsewhere. Complexity, evolution and classification suggest similar studies as in phylogenetics. Astrocladistics has opened the pathway.


\begin{thebibliography}{19}
\expandafter\ifx\csname natexlab\endcsname\relax\def\natexlab#1{#1}\fi
\expandafter\ifx\csname url\endcsname\relax
  \def\url#1{\texttt{#1}}\fi
\expandafter\ifx\csname urlprefix\endcsname\relax\def\urlprefix{URL }\fi
\providecommand{\eprint}[2][]{\url{#2}}
\providecommand{\bibinfo}[2]{#2}
\ifx\xfnm\relax \def\xfnm[#1]{\unskip,\space#1}\fi
\bibitem[{Bell(2005)}]{Bell2005}
\bibinfo{author}{Bell, E.}, \bibinfo{year}{2005}.
\newblock \bibinfo{title}{Galaxy Assembly. Planets to Cosmology: Essential
  Science in Hubble's Final Years}.
\newblock
\newblock \bibinfo{publisher}{Cambridge: CUP}. \eprint{astro-ph/0408023}.
\bibitem[{Buchanan and Collard(2008)}]{buchanan2008}
\bibinfo{author}{Buchanan, B.}, \bibinfo{author}{Collard, M.},
  \bibinfo{year}{2008}.
\newblock \bibinfo{title}{Phenetics, cladistics, and the search for the alaskan
  ancestors of the paleoindians: a reassessment of relationships among the
  clovis, nenana, and denali archaeological complexes}.
\newblock \textit{\bibinfo{journal}{Journal of Archaeological Science}}
\newblock \bibinfo{volume}{35}, \bibinfo{pages}{1683--1694}.
\bibitem[{Cabanac et~al.(2002)Cabanac, de~Lapparent and Hickson}]{cabanac2002}
\bibinfo{author}{Cabanac, R.A.}, \bibinfo{author}{de~Lapparent, A.},
  \bibinfo{author}{Hickson, P.}, \bibinfo{year}{2002}.
\newblock \textit{\bibinfo{journal}{Astronomy \& Astrophysics}}
\newblock \bibinfo{volume}{389}, \bibinfo{pages}{1090--1116}.
\bibitem[{Chattopadhyay et~al.(2009a)Chattopadhyay, Chattopadhyay, Davoust,
  Mondal and Sharina}]{Chattopadhyay2009a}
\bibinfo{author}{Chattopadhyay, A.}, \bibinfo{author}{Chattopadhyay, T.},
  \bibinfo{author}{Davoust, E.}, \bibinfo{author}{Mondal, S.},
  \bibinfo{author}{Sharina, M.}, \bibinfo{year}{2009}a.
\newblock \bibinfo{title}{Study of ngc 5128 globular clusters under
  multivariate statistical paradigm}.
\newblock \textit{\bibinfo{journal}{Astrophysical Journal}}
\newblock \bibinfo{volume}{705}, \bibinfo{pages}{1533}.
  \eprint{arXiv:0909.4161}.
\bibitem[{Chattopadhyay et~al.(2009b)Chattopadhyay, Babu, Chattopadhyay and
  Mondal}]{Chattopadhyay2009b}
\bibinfo{author}{Chattopadhyay, T.}, \bibinfo{author}{Babu, J.},
  \bibinfo{author}{Chattopadhyay, A.}, \bibinfo{author}{Mondal, S.},
  \bibinfo{year}{2009}b.
\newblock \bibinfo{title}{Horizontal branch morphology of globular clusters: A
  multivariate statistical analysis}.
\newblock \textit{\bibinfo{journal}{Astrophysical Journal}}
\newblock \bibinfo{volume}{700}, \bibinfo{pages}{1768}.
\bibitem[{Chattopadhyay and Chattopadhyay(2006)}]{Chattopadhyay2006}
\bibinfo{author}{Chattopadhyay, T.}, \bibinfo{author}{Chattopadhyay, A.},
  \bibinfo{year}{2006}.
\newblock \bibinfo{title}{Objective classification of spiral galaxies having
  extended rotation curves beyond the optical radius.}
\newblock \textit{\bibinfo{journal}{The Astronomical Journal}}
\newblock \bibinfo{volume}{131}, \bibinfo{pages}{2452–2468}.
\bibitem[{Chattopadhyay and Chattopadhyay(2007)}]{Chattopadhyay2007}
\bibinfo{author}{Chattopadhyay, T.}, \bibinfo{author}{Chattopadhyay, A.},
  \bibinfo{year}{2007}.
\newblock \bibinfo{title}{Globular clusters of local group – statistical
  analysis}.
\newblock \textit{\bibinfo{journal}{Astronomy \& Astrophysics}}
\newblock \bibinfo{volume}{472}, \bibinfo{pages}{131--140}.
\bibitem[{Chattopadhyay et~al.(2007)Chattopadhyay, Misra, Naskar and
  Chattopadhyay}]{ChattopadhyayGRB2007}
\bibinfo{author}{Chattopadhyay, T.}, \bibinfo{author}{Misra, R.},
  \bibinfo{author}{Naskar, M.}, \bibinfo{author}{Chattopadhyay, A.},
  \bibinfo{year}{2007}.
\newblock \bibinfo{title}{Statistical evidences of three classes of gamma ray
  bursts}.
\newblock \textit{\bibinfo{journal}{Astrophysical Journal}}
\newblock \bibinfo{volume}{667}, \bibinfo{pages}{1017}.
  \eprint{arXiv:0705.4020}.
\bibitem[{Chattopadhyay et~al.(2008)Chattopadhyay, Mondal and
  Chattopadhyay}]{Chattopadhyay2008}
\bibinfo{author}{Chattopadhyay, T.}, \bibinfo{author}{Mondal, S.},
  \bibinfo{author}{Chattopadhyay, A.}, \bibinfo{year}{2008}.
\newblock \bibinfo{title}{Globular clusters in the milky way and dwarf galaxies
  - a distribution-free statistical comparison}.
\newblock \textit{\bibinfo{journal}{Astrophysical Journal}}
\newblock \bibinfo{volume}{683}, \bibinfo{pages}{172}.
\bibitem[{{Fraix-Burnet}(2006)}]{DFB06}
\bibinfo{author}{{Fraix-Burnet}, D.}, \bibinfo{year}{2006}.
\newblock \bibinfo{title}{Determining the evolutionary history of galaxies by
  astrocladistics: some results on close galaxies}, in:
  \bibinfo{editor}{D.~Barret, F.~Casoli, S.C.F.C.T.C.},
  \bibinfo{editor}{Pagani, L.} (Eds.), \bibinfo{booktitle}{Journées de la
  SF2A, Paris (France)},
\newblock \bibinfo{publisher}{Société Française d'Astronomie et
  d'Astrophysique (SF2A)}.
  \eprint{http://hal.archives-ouvertes.fr/ccsd-00104352}.
\bibitem[{{Fraix-Burnet}(2009)}]{DFB09}
\bibinfo{author}{{Fraix-Burnet}, D.}, \bibinfo{year}{2009}.
\newblock \bibinfo{title}{Evolutionary Biology Concept, Modeling, and
  Application}. \bibinfo{publisher}{Springer Berlin Heidelberg}. chapter
  \bibinfo{chapter}{Galaxies and Cladistics}.
\newblock Biomedical and Life Sciences,
\newblock pp. \bibinfo{pages}{363--378}. \eprint{arXiv:0909.4164}.
\bibitem[{{Fraix-Burnet} et~al.(2006a){Fraix-Burnet}, {C}holer and
  {D}ouzery}]{FCD06}
\bibinfo{author}{{Fraix-Burnet}, D.}, \bibinfo{author}{{C}holer, P.},
  \bibinfo{author}{{D}ouzery, E.}, \bibinfo{year}{2006}a.
\newblock \bibinfo{title}{{T}owards a {P}hylogenetic {A}nalysis of {G}alaxy
  {E}volution : a {C}ase {S}tudy with the {D}warf {G}alaxies of the {L}ocal
  {G}roup}.
\newblock \textit{\bibinfo{journal}{{A}stronomy and {A}strophysics}}
\newblock \bibinfo{volume}{455}, \bibinfo{pages}{845--851}.
  \eprint{astro-ph/0605221}.
\bibitem[{{Fraix-Burnet} et~al.(2006b){Fraix-Burnet}, {C}holer, {D}ouzery and
  {V}erhamme}]{jc1}
\bibinfo{author}{{Fraix-Burnet}, D.}, \bibinfo{author}{{C}holer, P.},
  \bibinfo{author}{{D}ouzery, E.}, \bibinfo{author}{{V}erhamme, A.},
  \bibinfo{year}{2006}b.
\newblock \bibinfo{title}{{A}strocladistics: a phylogenetic analysis of galaxy
  evolution {I}. {C}haracter evolutions and galaxy histories}.
\newblock \textit{\bibinfo{journal}{{J}ournal of {C}lassification}}
\newblock \bibinfo{volume}{23}, \bibinfo{pages}{31--56}.
\bibitem[{{Fraix-Burnet} et~al.(2009){Fraix-Burnet}, {D}avoust and
  {C}harbonnel}]{FDC09}
\bibinfo{author}{{Fraix-Burnet}, D.}, \bibinfo{author}{{D}avoust, E.},
  \bibinfo{author}{{C}harbonnel, C.}, \bibinfo{year}{2009}.
\newblock \bibinfo{title}{{T}he environment of formation as a second parameter
  for globular cluster classification}.
\newblock \textit{\bibinfo{journal}{MNRAS}}
\newblock \bibinfo{volume}{398}, \bibinfo{pages}{1706--1714}.
  \eprint{arXiv:0906.3458}.
\bibitem[{{Fraix-Burnet} et~al.(2006c){Fraix-Burnet}, {D}ouzery, {C}holer and
  {V}erhamme}]{jc2}
\bibinfo{author}{{Fraix-Burnet}, D.}, \bibinfo{author}{{D}ouzery, E.},
  \bibinfo{author}{{C}holer, P.}, \bibinfo{author}{{V}erhamme, A.},
  \bibinfo{year}{2006}c.
\newblock \bibinfo{title}{{A}strocladistics: a phylogenetic analysis of galaxy
  evolution {II}. {F}ormation and diversification of galaxies}.
\newblock \textit{\bibinfo{journal}{{J}ournal of {C}lassification}}
\newblock \bibinfo{volume}{23}, \bibinfo{pages}{57--78}.
\bibitem[{{Fraix-Burnet} et~al.(2010){Fraix-Burnet}, Dugué, Chattopadhyay,
  Chattopadhyay and Davoust}]{Fraix2010}
\bibinfo{author}{{Fraix-Burnet}, D.}, \bibinfo{author}{Dugué, M.},
  \bibinfo{author}{Chattopadhyay, A.}, \bibinfo{author}{Chattopadhyay, T.},
  \bibinfo{author}{Davoust, E.}, \bibinfo{year}{2010}.
\newblock \bibinfo{title}{Structures in the fundamental plane of early-type
  galaxies}.
\newblock \textit{\bibinfo{journal}{Monthly Notices of the Royal Astronomical
  Society}}
\newblock \bibinfo{volume}{accepted for publication}.
  \eprint{http://fr.arxiv.org/abs/1005.5645}.
\bibitem[{Neistein et~al.(2006)Neistein, van~den Bosch and
  Dekel}]{Neistein2006}
\bibinfo{author}{Neistein, E.}, \bibinfo{author}{van~den Bosch, F.},
  \bibinfo{author}{Dekel, A.}, \bibinfo{year}{2006}.
\newblock \bibinfo{title}{Natural downsizing in hierarchical galaxy formation}.
\newblock \textit{\bibinfo{journal}{Monthly Notices of the Royal Astronomical
  Society}}
\newblock \bibinfo{volume}{372}, \bibinfo{pages}{933--948}.
  \eprint{astro-ph/0605045}.
\bibitem[{Recio-Blanco et~al.(2006)Recio-Blanco, Aparicio, Piotto, De~Angeli
  and Djorgovski}]{RecioBlanco2006}
\bibinfo{author}{Recio-Blanco, A.}, \bibinfo{author}{Aparicio, A.},
  \bibinfo{author}{Piotto, G.}, \bibinfo{author}{De~Angeli, F.},
  \bibinfo{author}{Djorgovski, S.}, \bibinfo{year}{2006}.
\newblock \bibinfo{title}{Multivariate analysis of globular cluster horizontal
  branch morphology: searching for the second parameter}.
\newblock \textit{\bibinfo{journal}{Astronomy \& Astrophysics}}
\newblock \bibinfo{volume}{452}.
  \eprint{http://arxiv.org/abs/astro-ph/0511704}.
\bibitem[{Wiley et~al.(1991)Wiley, Siegel-Causey, Brooks and Funk}]{Wiley1991}
\bibinfo{author}{Wiley, E.}, \bibinfo{author}{Siegel-Causey, D.},
  \bibinfo{author}{Brooks, D.}, \bibinfo{author}{Funk, V.},
  \bibinfo{year}{1991}.
\newblock \bibinfo{title}{The Compleat Cladist: A Primer of Phylogenetic
  Procedures}.
\newblock
\newblock \bibinfo{publisher}{The University of Kansas, Museum of Natural
  History, Special Publication No. 19}.

\end{thebibliography}

\end{document}